\documentclass[12pt]{article}
\usepackage{amsfonts}

\begin{document}
\title{Reconstruction
of density functions by $sk$-splines}
\author{
A. Kushpel and J. Levesley \\
Department of Mathematics\\
University of Leicester, UK \\
ak412@le.ac.uk, jl1@le.ac.uk
}
\date{April 18, 2014}

\maketitle

\begin{abstract}

Reconstruction of density functions and their characteristic functions by
radial basis functions with scattered data points is a popular topic in the
theory of pricing of basket options.

Such functions are usually entire in $\mathbb{C}^{n}$ or admit an analytic
extension into an appropriate tube in $\mathbb{C}^{n}$ and "bell-shaped"
with rapidly decaying tails.

Unfortunately, the domain of such functions (which are important in
practical applications) is not compact (e.g., $\mathbb{R}^{n})$ which
creates difficulties of a fundamental nature. Frequently used approach to
overcome this problem is an "appropriate" compactification (or truncation)
of the domain. Then in this compact domain we can try to interpolate (or
quasi-interpolate) by radial basis functions with a finite number of data
points. However, solution of the respective interpolation problem is
connected with inversion of "big" matrices and remains a challenging
computational problem. Moreover, the accuracy of approximation can not be
improved after truncation no matter how many data points we take in the
truncated domain. Also, the Fourier transform of truncated characteristic
function (which is a density function up to some multiplicative factor)
usually is not integrable which creates additional technical difficulties.
To avoid this range of problems many authors tried to construct extensions
of truncated characteristic functions into a larger but still compact domain
using algebraic polynomials and ingnoring their analytic smoothness. Of
course, such approach can not produce effective and saturation free
algorithms.

In this article we present a different approach. The values of a given
characteristic function can be computed on a rectangular mesh which allows us
to solve the respective interpolation problem explicitly under very general
conditions on the kernel function. Then we calculate explicitly Fourier
transform of such interpolant to obtain an approximant for the density
function.
\end{abstract}

\section{Introduction}

Consider a frictionless market with no arbitrage opportunities and
a constant riskless interest rate $r>0$. Let $S_{j,t}$, $1\leq j\leq n,t\geq
0$, be $n$ asset price processes. The common spread option with maturity $%
T>0 $ and strike $K\geq 0$ is the contract that pays $\left(
S_{1,T}-\sum_{j=2}^{n}S_{j,T}-K\right) _{+}$ at time $T$, where $\left(
a\right) _{+}:=\max \left\{ a,0\right\} $. There is a wide range of such
options traded across different sectors of financial markets. For instance,
the crack spread and crush spread options in the commodity markets \cite%
{Mbanefo}, \cite{Shimko}, credit spread options in the fixed income markets,
index spread options in the equity markets \cite{Duan} and the spark
(fuel/electricity) spread options in the energy markets \cite{Deng}, \cite%
{Pilipovic}.

Assuming the existence of a risk-neutral equivalent martingale measure $%
\mathbb{Q}$ we get the following pricing formula for the value at time $0$,
\[
V=e^{-rT}\mathbb{E}^{\mathbb{Q}}\left[ \varphi \right] ,
\]
where $\varphi $ is a reward function and the expectation is taken with
respect to the equivalent martingale measure. Usually, the reward function
has a simple structure, hence the main problem is to approximate properly
the respective density function and then to approximate $\mathbb{E}^{\mathbb{%
Q}}\left[ \varphi \right] $.

Let $\mathbf{x}=\left( x_{1},\cdot \cdot \cdot ,x_{n}\right) $ and $\mathbf{y%
}=\left( y_{1},\cdot \cdot \cdot ,y_{n}\right) $ be two vectors in $\mathbb{R%
}^{n}$, $\left\langle \mathbf{x,y}\right\rangle \mathbf{:=}%
\sum_{k=1}^{n}x_{k}y_{k}$ be the usual scalar product and $\left\vert
\mathbf{x}\right\vert ^{2}:=\left\Vert \mathbf{x}\right\Vert
_{2}:=\left\langle \mathbf{x,x}\right\rangle ^{1/2}$. For an integrable on $%
\mathbb{R}^{n}$ function, i.e., $f(\mathbf{x})\in L_{1}\left( \mathbb{R}%
^{n}\right) $ define its Fourier transform
\[
\mathbf{F}f(\mathbf{y})=\int_{\mathbb{R}^{n}}\exp \left( -i\left\langle
\mathbf{x,y}\right\rangle \right) f(\mathbf{x})d\mathbf{x}.
\]
and its formal inverse as%
\[
\left( \mathbf{F}^{-1}f\right) (\mathbf{x})=\frac{1}{\left( 2\pi \right) ^{n}%
}\int_{\mathbb{R}^{n}}\exp \left( -i\left\langle \mathbf{x,y}\right\rangle
\right) f(\mathbf{y})d\mathbf{y}.
\]
The characteristic function of the distribution of $\mathbf{X}_{t}$ of any L%
\'{e}vy process can be represented in the form
\[
\mathbb{E}^{\mathbb{Q}}\left[ \exp \left( \left\langle i\mathbf{x,X}%
_{t}\right\rangle \right) \right] =e^{-t\psi ^{\mathbb{Q}}\left( \mathbf{x}%
\right) }
\]
\[
=\left( 2\pi \right) ^{n}\mathbf{F}^{-1}p_{t}^{\mathbb{Q}}\left( \mathbf{x}%
\right) ,
\]
where $p_{t}^{\mathbb{Q}}\left( \mathbf{x}\right) $ is the density function
of $\mathbf{X}_{t}$, $\mathbf{x}\in \mathbb{R}^{n}$, $t\in \mathbb{R}_{+}$
and the function $\psi ^{\mathbb{Q}}\left( \mathbf{x}\right) $ \ is uniquely
determined. This function is called the characteristic exponent. Vice versa,
a L\'{e}vy process $\mathbf{X}=\{\mathbf{X}_{t}\}_{t\in \mathbb{R}_{+}}$ is
determined uniquely by its characteristic exponent $\psi ^{\mathbb{Q}}\left(
\mathbf{x}\right) $. In particular, density function $p_{t}^{\mathbb{Q}}$
can be expressed as
\[
p_{t}^{\mathbb{Q}}\left( \cdot \right) =\left( 2\pi \right) ^{-n}\int_{%
\mathbb{R}^{n}}\exp \left( -i\left\langle \cdot \mathbf{,x}\right\rangle
-t\psi ^{\mathbb{Q}}\left( \mathbf{x}\right) \right) d\mathbf{x}
\]
\[
=\left( 2\pi \right) ^{-n}\mathbf{F}\left( \exp \left( -t\psi ^{\mathbb{Q}%
}\left( \mathbf{x}\right) \right) \right) \left( \cdot \right)
\]
\[
=\left( 2\pi \right) ^{-n}\mathbf{F}\left( \Phi ^{\mathbb{Q}}\left( \mathbf{%
x,}t\right) \right) \left( \cdot \right) ,
\]
where $\Phi ^{\mathbb{Q}}\left( \mathbf{x,}t\right) $ is the characteristic
function of $\mathbf{X}=\{\mathbf{X}_{t}\}_{t\in \mathbb{R}_{+}}.$ Let $%
\Lambda :=\left\{ \mathbf{x}_{\mathbf{k}}\right\} $ be an additive group of
lattice points in $\mathbb{R}^{n}$ \ and $K\left( \cdot \right) $ be a fixed
kernel function (The reader should't mix the strike price $K$ with the
kernel function $K\left( \cdot \right) $). Assume that the interpolant
\[
sk\left( \Phi ^{\mathbb{Q}},\mathbf{x}\right) :=\sum_{\mathbf{\Lambda }}c_{%
\mathbf{k}}K\left( \mathbf{x-x}_{\mathbf{k}}\right)
\]
for $\Phi ^{\mathbb{Q}}\left( \mathbf{x,}t\right) $ exists and unique. Then,
formally, we get
\[
p_{t}^{\mathbb{Q}}\left( \cdot \right) \approx \left( 2\pi \right)
^{-n}\sum_{\mathbf{\Lambda }}c_{\mathbf{k}}\left( \Phi ^{\mathbb{Q}}\left(
\mathbf{x}_{\mathbf{k}}\mathbf{,}t\right) \right) \mathbf{F}\left( K\left(
\mathbf{x-x}_{\mathbf{k}}\right) \right) \left( \cdot \right)
\]
\[
=\left( 2\pi \right) ^{-n}\mathbf{F}\left( K\right) \left( \cdot \right)
\sum_{\mathbf{\Lambda }}c_{\mathbf{k}}\left( \Phi ^{\mathbb{Q}}\left(
\mathbf{x}_{\mathbf{k}}\mathbf{,}t\right) \right) \exp \left( i\left\langle
\cdot ,\mathbf{x}_{\mathbf{k}}\right\rangle \right) .
\]
In what follows we give an explicit form of $c_{\mathbf{k}}\left( \Phi ^{%
\mathbb{Q}}\left( \mathbf{x}_{\mathbf{k}}\mathbf{,}t\right) \right) $ and $%
\mathbf{F}\left( K\left( \mathbf{x-x}_{k}\right) \right) \left( \cdot
\right) $ which will give us an approximant for the density function $p_{t}^{%
\mathbb{Q}}$. Remark that in many important cases the coefficients $%
\left\vert c_{\mathbf{k}}\left( \Phi ^{\mathbb{Q}}\left( \mathbf{x}_{\mathbf{%
k}}\mathbf{,}t\right) \right) \right\vert $ decay exponentially fast as $%
\left\vert \mathbf{k}\right\vert \rightarrow \infty $.

Let us consider in more details the problem of interpolation in $\mathbb{R}%
^{n}$. Let $f$ be a continuous function on $\mathbb{R}^{n},$ $f\in C\left(
\mathbb{R}^{n}\right) $ and
\[
\sum_{\mathbf{x}_{j}\in \Lambda }c_{j}K\left( \mathbf{x}_{k}-\mathbf{x}%
_{j}\right) =f\left( \mathbf{x}_{k}\right) ,\mathbf{x}_{k},\mathbf{x}_{j}\in
\Lambda .
\]
Of course, it is very difficult (or in general impossible) to get an
explicit solution of the interpolation problem. But, if we assume some
regularity condition on the data points $\left\{ \mathbf{x}_{k}\right\} $ it
is still possible to solve the interpolation system. We \ give here an
explicit solution of the interpolation problem in the case of a uniform mesh
on $\mathbb{R}^{n}$.

Let $L_{p}(\mathbb{R}^{n})$ be the usual space of $p$-integrable functions
equipped with the norm
\[
\Vert f\Vert _{p}=\Vert f\Vert _{L_{p}(\mathbb{R}^{n})}:=\left\{
\begin{array}{cc}
\left( \int_{\mathbb{R}^{n}}\left\vert f(\mathbf{x})\right\vert ^{p}d\mathbf{%
x}\right) ^{1/p}, & 1\leq p<\infty , \\
\mathrm{ess}\,\,\sup_{\mathbf{x}\in \mathbb{R}^{n}}|f(\mathbf{x})|, &
p=\infty .%
\end{array}%
\right.
\]
To justify an inversion formula we will need celebrated Planchrel's theorem.

{\bf Theorem 1}
{\em (Plancherel) The Fourier transform is a linear continuous operator from $%
L_{2}\left( \mathbb{R}^{n}\right) $ onto $L_{2}\left( \mathbb{R}^{n}\right)
. $ The inverse Fourier transform, $\mathbf{F}^{-1},$ can be obtained by
letting
\[
\left( \mathbf{F}^{-1}g\right) \left( \mathbf{x}\right) =\frac{1}{(2\pi )^{n}%
}\left( \mathbf{F}g\right) \left( -\mathbf{x}\right)
\]
for any $g\in L_{2}\left( \mathbb{R}^{n}\right) .$}

Let us describe first the one-dimensional situation on $\mathbb{T}^{1}$. For
a given $m\in \mathbb{N}$ let $\Lambda _{m}=\{0=x_{0}<\cdots
<x_{n-1}<x_{m}=2\pi \}$ be an arbitrary partition of $[0,2\pi )$ and $K$ be
a continuous function. Then
\[
sk(x)=\sum_{x_{k}\in \Lambda _{m}}\,\,c_{k}K(x-x_{k}),\,\,\,c_{k}\in \mathbb{%
R}.
\]
Denote by $SK(\Lambda _{m})$ the space of $sk$-splines, i.e.
\[
SK(\Lambda _{m})=\mathrm{span}\{K(x-x_{m}),\,x_{k}\in \Lambda _{m}\}.
\]
See \cite{izv} for more information. Let $y\in \mathbb{R}$ be a fixed
parameter and $y_{k}=y+x_{k}$, $1\leq k\leq m$ be the points of
interpolation. If the interpolation problem has a unique solution then the
spline interpolant $sk(x,y,\Lambda _{m},f)=sk(x,f)$ with knots $x_{k}$, $%
1\leq k\leq m$ and points of interpolation $y_{k}$ can be written in the
form
\[
sk(x)=\sum_{k\in \Lambda _{m}}\,f(y_{k})\tilde{sk}_{k}(x),
\]
where
\[
\tilde{sk}_{k}(y_{s})=\left\{
\begin{array}{cc}
1, & k=s, \\
0, & k\neq s.%
\end{array}%
\right.
\]
are the fundamental $sk$-splines. It is important in various applications to
have an explicit form of the Fourier series expansions for the fundamental $%
sk$-splines. As a motivating example consider $sk$-splines on the uniform
greed $\Lambda _{m}=\{x_{k}=2\pi /m,\,1\leq k\leq m\}$, with the points of
interpolation ${y_{1},\cdots ,y_{m}}$ where $y\in \mathbb{R}$ is a fixed
parameter,
\[
sk(x)=c_{0}+\sum_{k=1}^{m}c_{k}K(x-x_{k}),\,\,\,\sum_{k=1}^{m}c_{k}=0,\,%
\,c_{k}\in \mathbb{R},\,\,1\leq k\leq m,
\]
In this case fundamental splines are just the shifts of $\tilde{sk}(\cdot )$%
, where
\[
\tilde{sk}(y_{k})=\left\{
\begin{array}{cc}
1, & k\equiv 0(\mathrm{mod}n), \\
0, & otherwise.
\end{array}%
\right.
\]
Let, in particular,
\[
K(x)=D_{r}(x)=\sum_{k=1}^{\infty }\frac{1}{k^{r}}\cos \left( kx+\frac{r\pi }{%
2}\right) ,\,\,r\in \mathbb{N}
\]
be the Bernoulli monospline. Then the space $SK(\Lambda _{m})$ is the space
of polynomial splines of order $r-1$, defect $1$with knots $x_{k}$, $1\leq
k\leq m$ and points of interpolation $y_{k}$, $1\leq k\leq m$.

First Fourier series expansions of fundamental splines have been obtained by
Golomb \cite{golomb} in the case $r=4$, and $y=0$, i.e. in the case of cubic
splines. It was shown that
\[
\tilde{sk}(x,0)=\tilde{sk}(x)=\frac{1}{m}+\frac{1}{m}\sum_{j=1}^{m-1}\frac{%
\rho _{j}(x)}{\rho _{j}(0)},
\]
where
\[
\rho _{j}(x)=\sum_{\nu =1}^{m}\cos \left( \frac{2\pi \nu j}{m}\right)
\,D_{4}\left( x-\frac{2\pi \nu }{m}\right) .
\]
In the case of a general kernel function $K\in C\left( \mathbb{T}^{1}\right)
$ and an arbitrary $y\in \mathbb{R}$ the respective results were established
in \cite{izv}. Namely, it was shown that
\[
\tilde{sk}(x,y)=\tilde{sk}(x)=\frac{1}{m}+\frac{1}{m}\sum_{j=1}^{m-1}\frac{%
\rho _{j}(x)\rho _{j}(y)+\sigma _{j}(x)\sigma _{j}(y)}{\rho
_{j}^{2}(y)+\sigma _{j}^{2}(y)},
\]
where
\[
\rho _{j}(x)=\sum_{\nu =1}^{m}\cos \left( \frac{2\pi \nu j}{m}\right)
\,K\left( x-\frac{2\pi \nu }{m}\right) ,\sigma _{j}(x)=\sum_{\nu =1}^{m}\sin
\left( \frac{2\pi \nu j}{m}\right) \,K\left( x-\frac{2\pi \nu }{m}\right) .
\]
Of course, to guarantee existence of fundamental splines for a given $y$ we
need to assume that $\max \{\rho _{j}^{2}(y),\sigma _{j}^{2}(y),1\leq j\leq
m\}>0$. A detailed study of such kind of conditions in terms of Fourier
coefficients of the kernel function $K$ can be found in \cite{izv}, \cite%
{skk} 
Different analogs of these results in multidimensional settings, on $\mathbb{%
T}^{d}$can be found in \cite{kushpel8}, \cite{kll}, \cite{kushpel9}. Remark
that the problem of convergence of $sk$-spline interpolants and
quasi-interpolants was considered in \cite{kushpel1}, \cite{kgh}, \cite%
{kushpel9}, \cite{kushpel-b1}, \cite{kushpel-b2} where it was shown that the
rate of convergence of $sk$-splines has the same order as the respective $n$%
-widths.

The main aim of this article is to establish representations of cardinal $sk$%
-splines on a uniform mesh in $\mathbb{R}^{n}$ and to apply these results to
the problem of recovery of density functions which are important in the
theory of pricing.

\section{Interpolation by sk-splines on $\mathbb{R}^{n}$}

Let $\mathbf{a}=\left( a_{1},\cdot \cdot \cdot ,a_{n}\right) $, $a_{k}>0$, $%
1\leq k\leq n$ be a fixed mesh parameter $\mathbf{m}=\left( m_{1},\cdot
\cdot \cdot ,m_{n}\right) \in \mathbb{Z}^{n}$ and
\[
\Omega _{\mathbf{a}}:=\left\{ \left( a_{1}m_{1},\cdot \cdot \cdot
,a_{n}m_{n}\right) \mathbf{|m\in }\mathbb{Z}^{n}\right\} \subset \mathbb{R}%
^{n}
\]
be a mesh in $\mathbb{R}^{n}$. \ Let $\mathbf{A}:=\mathrm{diag}\left(
a_{1},\cdot \cdot \cdot ,a_{n}\right) $, then the mesh points are $\mathbf{x}%
_{\mathbf{m}}:=\mathbf{Am}^{T}$. \ For a fixed continuous kernel function $K$%
, the space $SK\left( \Omega _{\mathbf{a}}\right) $ of $sk$-splines on $%
\Omega _{\mathbf{a}}$ is the space of functions representable in the form
\[
sk\left( \mathbf{x}\right) =\sum_{\mathbf{m\in }\mathbb{Z}^{n}}c_{\mathbf{m}%
}K\left( \mathbf{x-x}_{\mathbf{m}}\right) ,
\]
where $c_{\mathbf{m}}\in \mathbb{R}$. Let $f\left( \mathbf{x}\right) $ be a
continuous function, $f:\mathbb{R}^{n}\rightarrow \mathbb{R}$. Consider the
problem of interpolation%
\[
sk\left( \mathbf{x}_{\mathbf{m}}\right) =f\left( \mathbf{x}_{\mathbf{m}%
}\right) ,
\]
where
\[
sk\left( \mathbf{x}_{\mathbf{m}}\right) =\sum_{\mathbf{m\in }\mathbb{Z}%
^{n}}c_{\mathbf{m}}^{\ast }K\left( \mathbf{x-x}_{\mathbf{m}}\right) ,c_{%
\mathbf{m}}^{\ast }\in \mathbb{R}.
\]
Even in the one-dimensional case the problem of interpolation not always has
a solution. If the solution exists then $sk$-spline interpolant can be
written in the form
\[
sk\left( \mathbf{x}\right) =\sum_{\mathbf{m\in }\mathbb{Z}^{n}}f\left(
\mathbf{x}_{\mathbf{m}}\right) \widetilde{sk}\left( \mathbf{x}-\mathbf{x}_{%
\mathbf{m}}\right) ,
\]
where $\widetilde{sk}\left( \mathbf{x}-\mathbf{x}_{\mathbf{m}}\right) $ are
fundamental $sk$-splines, i.e.%
\[
\widetilde{sk}\left( \mathbf{x}_{\mathbf{m}}\right) =\left\{
\begin{array}{cc}
1, & \mathbf{m=0,} \\
0, & \mathbf{m\neq 0.}%
\end{array}%
\right.
\]

{\bf Theorem 2}
\begin{em}
\label{cardinal sk-spline} Let $K:\mathbb{R}^{n}\rightarrow \mathbb{R}$ be
such that $K\in L_{2}\left( \mathbb{R}^{n}\right) \cap C\left( \mathbb{R}%
^{n}\right) ,$
\[
\sum_{\mathbf{m}\in \mathbb{Z}^{n}}\mathbf{F}\left( K\right) \left( \mathbf{%
z+}2\pi \mathbf{A}^{-1}\mathbf{m}^{T}\right) \neq 0,\forall \mathbf{z\in }%
2\pi \mathbf{Q}_{\mathbf{a}},
\]
where $\mathbf{Q}_{\mathbf{a}}:=\left\{ \mathbf{x|x=}\left( x_{1},\cdot
\cdot \cdot ,x_{n}\right) \mathbf{\in }\mathbb{R}^{n},0\leq x_{k}\leq
1/a_{k},1\leq k\leq n\right\} ,$ and the function
\[
\frac{1}{\sum_{\mathbf{m}\in \mathbb{Z}^{n}}\mathbf{F}\left( K\right) \left(
\mathbf{z+}2\pi \mathbf{A}^{-1}\mathbf{m}^{T}\right) }
\]
can be represented by its Fourier series, i.e. for any $\mathbf{z\in }%
\mathbb{R}^{n}$, \
\[
\frac{1}{\sum_{\mathbf{m}\in \mathbb{Z}^{n}}\mathbf{F}\left( K\right) \left(
\mathbf{z+}2\pi \mathbf{A}^{-1}\mathbf{m}^{T}\right) }=\sum_{\mathbf{s}\in
\mathbb{Z}^{n}}\alpha _{\mathbf{s}}\exp \left( -i\left\langle \mathbf{As}%
^{T},\mathbf{z}\right\rangle \right)
\]
then%
\[
\widetilde{sk}\left( \mathbf{x}\right) =\frac{\det \left( A\right) }{\left(
2\pi \right) ^{n}}\int_{\mathbb{R}^{n}}\frac{\mathbf{F}\left( K\right)
\left( \mathbf{z}\right) }{\sum_{\mathbf{m}\in \mathbb{Z}^{n}}\mathbf{F}%
\left( K\right) \left( \mathbf{z+}2\pi \mathbf{A}^{-1}\mathbf{m}^{T}\right) }%
\exp \left( i\left\langle \mathbf{z},\mathbf{x}\right\rangle \right) d%
\mathbf{z}
\]
and this representation is unique.
\end{em}

{\bf Proof}
Since%
\[
\frac{1}{\sum_{\mathbf{m}\in \mathbb{Z}^{n}}\mathbf{F}\left( K\right) \left(
\mathbf{z+}2\pi \mathbf{A}^{-1}\mathbf{m}^{T}\right) }=\sum_{\mathbf{s}\in
\mathbb{Z}^{n}}\alpha _{\mathbf{s}}\exp \left( -i\left\langle \mathbf{As}%
^{T},\mathbf{z}\right\rangle \right)
\]
then
\[
\widetilde{sk}\left( \mathbf{x}\right) =\frac{\det \left( A\right) }{\left(
2\pi \right) ^{n}}\int_{\mathbb{R}^{n}}\frac{\mathbf{F}\left( K\right)
\left( \mathbf{z}\right) }{\sum_{\mathbf{m}\in \mathbb{Z}^{n}}\mathbf{F}%
\left( K\right) \left( \mathbf{z+}2\pi \mathbf{A}^{-1}\mathbf{m}^{T}\right) }%
\exp \left( i\left\langle \mathbf{x},\mathbf{z}\right\rangle \right) d%
\mathbf{z}
\]
\[
=\frac{\det \left( A\right) }{\left( 2\pi \right) ^{n}}\int_{\mathbb{R}^{n}}%
\mathbf{F}\left( K\right) \left( \mathbf{z}\right) \left( \sum_{\mathbf{s}%
\in \mathbb{Z}^{n}}\alpha _{\mathbf{s}}\exp \left( -i\left\langle \mathbf{As}%
^{T},\mathbf{z}\right\rangle \right) \right) \exp \left( i\left\langle
\mathbf{x,z}\right\rangle \right) d\mathbf{z}
\]
\[
=\frac{\det \left( A\right) }{\left( 2\pi \right) ^{n}}\sum_{\mathbf{s}\in
\mathbb{Z}^{n}}\alpha _{\mathbf{s}}\int_{\mathbb{R}^{n}}\mathbf{F}\left(
K\right) \left( \mathbf{z}\right) \exp \left( i\left\langle \mathbf{x}-%
\mathbf{As}^{T},\mathbf{z}\right\rangle \right) d\mathbf{z.}
\]
Since $K\in L_{2}\left( \mathbb{R}^{n}\right) $ then by Plancherel's theorem
\[
\widetilde{sk}\left( \mathbf{x}\right) =\det \left( A\right) \sum_{\mathbf{s}%
\in \mathbb{Z}^{n}}\alpha _{\mathbf{s}}K\left( \mathbf{x}-\mathbf{As}%
^{T}\right) ,
\]
so that $\widetilde{sk}\left( \mathbf{x}\right) \in SK\left( \Omega _{%
\mathbf{a}}\right) $. Hence%
\[
\widetilde{sk}\left( \mathbf{Am}^{T}\right) =\frac{\det \left( A\right) }{%
\left( 2\pi \right) ^{n}}\int_{\mathbb{R}^{n}}\frac{\mathbf{F}\left(
K\right) \left( \mathbf{z}\right) }{\sum_{\mathbf{m}\in \mathbb{Z}^{n}}%
\mathbf{F}\left( K\right) \left( \mathbf{z+}2\pi \mathbf{A}^{-1}\mathbf{m}%
^{T}\right) }\exp \left( i\left\langle \mathbf{z},\mathbf{Am}%
^{T}\right\rangle \right) d\mathbf{z}
\]
\[
=\frac{\det \left( A\right) }{\left( 2\pi \right) ^{n}}\sum_{\mathbf{l}\in
\mathbb{Z}^{n}}\int_{2\pi \mathbf{A}^{-1}\mathbf{l}^{T}+2\pi \mathbf{Q}_{%
\mathbf{a}}}\frac{\mathbf{F}\left( K\right) \left( \mathbf{z}\right) }{\sum_{%
\mathbf{m}\in \mathbb{Z}^{n}}\mathbf{F}\left( K\right) \left( \mathbf{z+}%
2\pi \mathbf{A}^{-1}\mathbf{m}^{T}\right) }\exp \left( i\left\langle \mathbf{%
z},\mathbf{Am}^{T}\right\rangle \right) d\mathbf{z}
\]
\[
=\frac{\det \left( A\right) }{\left( 2\pi \right) ^{n}}
\]
\[
\times \sum_{\mathbf{l}\in \mathbb{Z}^{n}}\int_{2\pi \mathbf{Q}_{\mathbf{a}}}%
\frac{\mathbf{F}\left( K\right) \left( \mathbf{z+}2\pi \mathbf{\mathbf{A}}%
^{-1}\mathbf{\mathbf{l}^{T}}\right) }{\sum_{\mathbf{m}\in \mathbb{Z}^{n}}%
\mathbf{F}\left( K\right) \left( \mathbf{z+}2\pi \mathbf{\mathbf{A}}^{-1}%
\mathbf{\mathbf{l}^{T}+}2\pi \mathbf{A}^{-1}\mathbf{m}^{T}\right) }\exp
\left( i\left\langle \mathbf{z+}2\pi \mathbf{\mathbf{A}}^{-1}\mathbf{\mathbf{%
l}^{T}},\mathbf{Am}^{T}\right\rangle \right) d\mathbf{z}
\]
\[
=\frac{\det \left( A\right) }{\left( 2\pi \right) ^{n}}\sum_{\mathbf{l}\in
\mathbb{Z}^{n}}\int_{2\pi \mathbf{Q}_{\mathbf{a}}}\frac{\mathbf{F}\left(
K\right) \left( \mathbf{z+}2\pi \mathbf{\mathbf{A}}^{-1}\mathbf{\mathbf{l}%
^{T}}\right) }{\sum_{\mathbf{m}\in \mathbb{Z}^{n}}\mathbf{F}\left( K\right)
\left( \mathbf{z+}2\pi \mathbf{A}^{-1}\mathbf{m}^{T}\right) }\exp \left(
i\left\langle \mathbf{z+}2\pi \mathbf{\mathbf{A}}^{-1}\mathbf{\mathbf{l}^{T}}%
,\mathbf{Am}^{T}\right\rangle \right) d\mathbf{z}
\]
\[
=\frac{\det \left( A\right) }{\left( 2\pi \right) ^{n}}\int_{2\pi \mathbf{Q}%
_{\mathbf{a}}}\frac{\sum_{\mathbf{l}\in \mathbb{Z}^{n}}\mathbf{F}\left(
K\right) \left( \mathbf{z+}2\pi \mathbf{\mathbf{A}}^{-1}\mathbf{\mathbf{l}%
^{T}}\right) }{\sum_{\mathbf{m}\in \mathbb{Z}^{n}}\mathbf{F}\left( K\right)
\left( \mathbf{z+}2\pi \mathbf{A}^{-1}\mathbf{m}^{T}\right) }\exp \left(
i\left\langle \mathbf{z+}2\pi \mathbf{\mathbf{A}}^{-1}\mathbf{\mathbf{l}^{T}}%
,\mathbf{Am}^{T}\right\rangle \right) d\mathbf{z}
\]
\[
=\frac{\det \left( A\right) }{\left( 2\pi \right) ^{n}}\int_{2\pi \mathbf{Q}%
_{\mathbf{a}}}\exp \left( i\left\langle \mathbf{z+}2\pi \mathbf{\mathbf{A}}%
^{-1}\mathbf{\mathbf{l}^{T}},\mathbf{Am}^{T}\right\rangle \right) d\mathbf{z}
\]
\[
=\frac{\det \left( A\right) }{\left( 2\pi \right) ^{n}}\int_{2\pi \mathbf{Q}%
_{\mathbf{a}}}\exp \left( i\left\langle \mathbf{z},\mathbf{Am}%
^{T}\right\rangle +2\pi i\left\langle \mathbf{\mathbf{l}},\mathbf{m}%
\right\rangle \right) d\mathbf{z}
\]
\[
=\frac{\det \left( A\right) }{\left( 2\pi \right) ^{n}}\int_{2\pi \mathbf{Q}%
_{\mathbf{a}}}\exp \left( i\left\langle \mathbf{z},\mathbf{Am}%
^{T}\right\rangle \right) d\mathbf{z}
\]
\[
=\frac{\det \left( A\right) }{\left( 2\pi \right) ^{n}}\int_{2\pi \mathbf{Q}%
_{\mathbf{a}}}\exp \left( i\sum_{k=1}^{n}a_{k}m_{k}z_{k}\right) d\mathbf{z}
\]
\[
=\frac{\det \left( A\right) }{\left( 2\pi \right) ^{n}}\prod%
\limits_{k=1}^{n}\int_{0}^{2\pi /a_{k}}\exp \left( ia_{k}m_{k}z_{k}\right)
dz_{k}
\]
\[
=\left\{
\begin{array}{cc}
1, & m_{k}=0,1\leq k\leq n, \\
0, & otherwise.
\end{array}%
\right. .
\]
Finally, we need to show that the representation of fundamental $sk-$spline
is unique. It is sufficient to show that the functions $\widetilde{sk}\left(
\mathbf{x}-\mathbf{x}_{\mathbf{m}}\right) ,\mathbf{x}_{\mathbf{m}}\in \Omega
_{\mathbf{a}}$ are linearly independent. Let $a_{\mathbf{m}}\in \mathbb{R},%
\mathbf{m\in }\mathbb{Z}^{n}$ be such that not all $a_{\mathbf{m}}$ are
zero. Let, in particular, $a_{\mathbf{s}}\neq 0$ for some $\mathbf{s\in }%
\mathbb{Z}^{n}.$ Consider a linear operator,
\[
\begin{array}{c}
A:C(\mathbb{R}^{n})\longrightarrow \mathbb{R} \\
f\left( \cdot \right) \longmapsto f\left( x_{\mathbf{s}}\right) .%
\end{array}%
\]
Assume that $\sum_{\mathbf{m\in }\mathbb{Z}^{n}}a_{\mathbf{m}}\widetilde{sk}%
\left( \mathbf{x}-\mathbf{x}_{\mathbf{m}}\right) \equiv 0$ then
\[
0=A\left( \sum_{\mathbf{m\in }\mathbb{Z}^{n}}a_{\mathbf{m}}\widetilde{sk}%
\left( \mathbf{x}-\mathbf{x}_{\mathbf{m}}\right) \right) =a_{\mathbf{s}%
}A\left( \widetilde{sk}\left( \mathbf{x}-\mathbf{x}_{\mathbf{s}}\right)
\right) =a_{\mathbf{s}},
\]
which is a contradiction. $\Box$

{\bf Theorem 3}
\begin{em}
In the assumptions of Theorem \ref{cardinal sk-spline}%
\[
p_{t}^{\mathbb{Q}}\left( \cdot \right) \approx \left( \frac{\det \left(
A\right) \mathbf{F}\left( K\right) \left( \mathbf{\cdot }\right) }{\left(
2\pi \right) ^{n}\sum_{\mathbf{m}\in \mathbb{Z}^{n}}\mathbf{F}\left(
K\right) \left( \cdot \mathbf{+}2\pi \mathbf{A}^{-1}\mathbf{m}^{T}\right) }%
\right) \left( \sum_{\mathbf{m}\in \mathbb{Z}^{n}}\Phi ^{\mathbb{Q}}\left(
\mathbf{x}_{\mathbf{m}}\mathbf{,}t\right) \exp \left( i\left\langle \cdot ,%
\mathbf{x}_{\mathbf{m}}\right\rangle \right) \left( \mathbf{\cdot }\right)
\right) ,
\]
as $\max \left\{ a_{k},1\leq k\leq n\right\} \rightarrow 0.$
\end{em}

{\bf Proof}
Observe that
\[
\widetilde{sk}\left( \mathbf{x}\right) \mathbf{=}\frac{\det \left( A\right)
}{\left( 2\pi \right) ^{n}}\int_{\mathbb{R}^{n}}\frac{\mathbf{F}\left(
K\right) \left( \mathbf{z}\right) }{\sum_{\mathbf{m}\in \mathbb{Z}^{n}}%
\mathbf{F}\left( K\right) \left( \mathbf{z+}2\pi \mathbf{A}^{-1}\mathbf{m}%
^{T}\right) }\exp \left( i\left\langle \mathbf{z},\mathbf{x}\right\rangle
\right) d\mathbf{z}
\]
\[
\mathbf{=}\det \left( A\right) \mathbf{F}^{-1}\left( \frac{\mathbf{F}\left(
K\right) \left( \mathbf{z}\right) }{\sum_{\mathbf{m}\in \mathbb{Z}^{n}}%
\mathbf{F}\left( K\right) \left( \mathbf{z+}2\pi \mathbf{A}^{-1}\mathbf{m}%
^{T}\right) }\right) \left( \mathbf{x}\right) ,\forall \mathbf{x}_{\mathbf{s}%
}\in \Omega _{\mathbf{a}}.
\]
Hence%
\[
p_{t}^{\mathbb{Q}}\left( \cdot \right) =\left( 2\pi \right) ^{-n}\mathbf{F}%
\left( \Phi ^{\mathbb{Q}}\left( \mathbf{x,}t\right) \right) \left( \cdot
\right)
\]
\[
\approx \left( 2\pi \right) ^{-n}\mathbf{F}\left( \sum_{\mathbf{m}\in
\mathbb{Z}^{n}}\Phi ^{\mathbb{Q}}\left( \mathbf{x}_{\mathbf{m}}\mathbf{,}%
t\right) \widetilde{sk}\left( \mathbf{x-x}_{\mathbf{m}}\right) \right)
\left( \cdot \right)
\]
\[
=\frac{\det \left( A\right) }{\left( 2\pi \right) ^{n}}\mathbf{F}\left(
\sum_{\mathbf{m}\in \mathbb{Z}^{n}}\Phi ^{\mathbb{Q}}\left( \mathbf{x}_{%
\mathbf{m}}\mathbf{,}t\right) \mathbf{F}^{-1}\left( \frac{\mathbf{F}\left(
K\right) \left( \mathbf{z}\right) }{\sum_{\mathbf{m}\in \mathbb{Z}^{n}}%
\mathbf{F}\left( K\right) \left( \mathbf{z+}2\pi \mathbf{A}^{-1}\mathbf{m}%
^{T}\right) }\right) \left( \mathbf{x-x}_{\mathbf{m}}\right) \right) \left(
\cdot \right)
\]
\[
=\frac{\det \left( A\right) }{\left( 2\pi \right) ^{n}}\left( \frac{\mathbf{F%
}\left( K\right) \left( \mathbf{\cdot }\right) }{\sum_{\mathbf{m}\in \mathbb{%
Z}^{n}}\mathbf{F}\left( K\right) \left( \cdot \mathbf{+}2\pi \mathbf{A}^{-1}%
\mathbf{m}^{T}\right) }\right) \left( \sum_{\mathbf{m}\in \mathbb{Z}%
^{n}}\Phi ^{\mathbb{Q}}\left( \mathbf{x}_{\mathbf{m}}\mathbf{,}t\right) \exp
\left( i\left\langle \cdot ,\mathbf{x}_{\mathbf{m}}\right\rangle \right)
\right) ,
\]
since%
\[
\frac{\mathbf{F}\left( K\right) \left( \mathbf{z}\right) }{\sum_{\mathbf{m}%
\in \mathbb{Z}^{n}}\mathbf{F}\left( K\right) \left( \mathbf{z+}2\pi \mathbf{A%
}^{-1}\mathbf{m}^{T}\right) }\in L_{2}\left( \mathbb{R}^{n}\right) .
\]
$\Box$

{\bf Example 1}
\begin{em}
Let $K\left( \mathbf{x}\right) $ be a Gaussian of the form
\[
K\left( \mathbf{x}\right) =K\left( x_{1},\cdot \cdot \cdot ,x_{n}\right)
=\exp \left( -\sum_{k=1}^{n}b_{k}x_{k}^{2}\right)
\]
and $B=diag\left( b_{1},\cdot \cdot \cdot ,b_{n}\right) $, then
\[
\mathbf{F}\left( K\right) \left( \mathbf{z}\right) =\pi ^{n/2}\left( \det
\left( B\right) \right) ^{-1/2}\exp \left( -\sum_{k=1}^{n}\frac{z_{k}^{2}}{%
4b_{k}}\right) .
\]
Applying Poisson summation formula and Plancherel's theorem we get%
\[
\sum_{\mathbf{m}\in \mathbb{Z}^{n}}\mathbf{F}\left( K\right) \left( \mathbf{%
z+}2\pi \mathbf{A}^{-1}\mathbf{m}^{T}\right)
\]
\[
=\sum_{\mathbf{m}\in \mathbb{Z}^{n}}\prod\limits_{k=1}^{n}\left( \mathbf{F}%
\exp \left( -b_{k}y_{k}^{2}\right) \right) \left( z_{k}+\frac{2\pi m_{k}}{%
a_{k}}\right)
\]
\[
=\prod\limits_{k=1}^{n}\sum_{m_{k}\in \mathbb{Z}}\left( \mathbf{F}\exp
\left( -b_{k}y_{k}^{2}\right) \right) \left( z_{k}+\frac{2\pi m_{k}}{a_{k}}%
\right)
\]
\[
=\prod\limits_{k=1}^{n}\left( \frac{a_{k}}{2\pi }\right) \sum_{m_{k}\in
\mathbb{Z}}\exp \left( ia_{k}m_{k}z_{k}\right) \mathbf{F\circ F}\left( \exp
\left( -b_{k}\left( \frac{a_{k}m_{k}}{2\pi }\right) ^{2}\right) \right)
\]
\[
=\prod\limits_{k=1}^{n}\left( \frac{a_{k}}{2\pi }\right) \left( 2\pi
\right) \sum_{m_{k}\in \mathbb{Z}}\exp \left( ia_{k}m_{k}z_{k}\right) \exp
\left( -b_{k}\left( \frac{a_{k}m_{k}}{2\pi }\right) ^{2}\right)
\]
\[
=\det \left( A\right) \prod\limits_{k=1}^{n}\sum_{m_{k}\in \mathbb{Z}}\exp
\left( ia_{k}m_{k}z_{k}\right) \exp \left( -b_{k}\left( \frac{a_{k}m_{k}}{%
2\pi }\right) ^{2}\right) .
\]
Hence, in this case%
\[
\frac{\det \left( A\right) \mathbf{F}\left( K\right) \left( \mathbf{z}%
\right) }{\left( 2\pi \right) ^{n}\sum_{\mathbf{m}\in \mathbb{Z}^{n}}\mathbf{%
F}\left( K\right) \left( \mathbf{z+}2\pi \mathbf{A}^{-1}\mathbf{m}%
^{T}\right) }
\]
\[
=\frac{\pi ^{n/2}\left( \det \left( B\right) \right) ^{-1/2}\exp \left(
-\sum_{k=1}^{n}\frac{z_{k}^{2}}{4b_{k}}\right) }{\left( 2\pi \right)
^{n}\prod\limits_{k=1}^{n}\sum_{m_{k}\in \mathbb{Z}}\exp \left(
ia_{k}m_{k}z_{k}\right) \exp \left( -b_{k}\left( \frac{a_{k}m_{k}}{2\pi }%
\right) ^{2}\right) }.
\]
\end{em}
\bigskip

\bigskip


\end{document}